\documentclass{llncs}
\usepackage[utf8]{inputenc}
\usepackage[english]{babel}
\usepackage{hyperref}
\usepackage{wrapfig}
\usepackage{proof}

\newcommand{\liquidsoap}{Liquidsoap}
\newcommand{\savonet}{Savonet}
\newcommand{\eg}{\emph{e.g.},}
\newcommand{\ie}{\emph{i.e.},}
\newcommand{\cf}{{cf.~}}
\newcommand{\TODO}[1]{}
\newcommand{\ignore}[1]{}
\newcommand{\fcaption}[1]{\vspace{-3ex}\caption{#1}\vspace{-4ex}}

\newcommand{\sub}{<:}

\usepackage{graphicx}
\usepackage[matrix,arrow,frame]{xy}

\usepackage{tikz}
\usetikzlibrary{snakes}

\title{\liquidsoap{}:\\
  a High-Level Programming~Language\\
  for Multimedia~Streaming}
\author{David Baelde\inst{1} \and Romain Beauxis\inst{2} \and Samuel Mimram\inst{3}}
\institute{
  University of Minnesota, USA
  \and
  Department of Mathematics, Tulane University, USA
  \and
  CEA LIST -- LMeASI, France
}

\hypersetup{
  pdftitle={\csname @title\endcsname},
  pdfauthor={David Baelde, Romain Beauxis and Samuel Mimram},
  unicode=true,
  colorlinks=true,
  linkcolor=black,
  citecolor=black,
  urlcolor=black
}

\begin{document}
\maketitle


\begin{abstract}
  Generating multimedia streams, such as in a netradio, is a task which is
  complex and difficult to adapt to every users' needs. We introduce a novel
  approach in order to achieve it, based on a dedicated high-level functional
  programming language, called \emph{\liquidsoap{}}, for generating,
  manipulating and broadcasting multimedia streams. Unlike traditional
  approaches, which are based on configuration files or static graphical
  interfaces, it also allows the user to build complex and highly customized
  systems. This language is based on a model for streams and contains operators
  and constructions, which make it adapted to the generation of streams. The
  interpreter of the language also ensures many properties concerning the good
  execution of the stream generation.
\end{abstract}

The widespread adoption of broadband internet in the last decades has
changed a lot our way of producing and consuming information. Classical
devices from the analog era, such as television or radio broadcasting devices
have been rapidly adapted to the digital world in order to benefit from the new
technologies available. While analog devices were mostly based on hardware
implementations, their digital counterparts often consist in software
implementations,
which potentially offers much more flexibility and modularity in their design.
However, there is still much progress to be done to unleash this
potential in many areas where software implementations remain pretty much as
hard-wired as their digital counterparts.

The design of domain specific languages is a powerful way of addressing
that challenge. It consists in identifying (or designing) relevant
domain-specific abstractions (construct well-behaved objects equipped
with enough operations) and make them available through a programming
language.
The possibility to manipulate rich high-level abstractions
by means of a flexible language
can often release creativity in unexpected ways.
To achieve this, a domain-specific language
should follow three fundamental principles. It should be
\begin{enumerate}
\item \emph{adapted}: users should be able to perform the required
  tasks in the domain of application of the language;
\item \emph{simple}: users should be able to perform the tasks in a simple way
  (this means that the language should be reasonably concise, but also
   understandable by users who might not be programming language experts);
\item \emph{safe}: the language should perform automatic checks
  to prevent as many errors as possible, using static analysis when
  possible.
\end{enumerate}
Balancing those requirements can be very difficult. This is perhaps
the reason why domain specific languages are not seen more often.
Another reason is that advanced concepts from the theory of
programming language and type systems
are often required to obtain a satisfying design.

In this paper,
we are specifically interested in the generation of multimedia streams,
notably containing audio and video.
Our primary targets are netradios, which continuously broadcast
audio streams to listeners over Internet.
At first, generating such a stream might seem to simply consist
in concatenating audio files. In practice, the needs of radio makers
are much higher than this.
For instance, a radio stream will often contain
commercial or informative jingles, which may be scheduled at regular
intervals, sometimes in between songs and sometimes mixed on top of them.
Also, a radio program may be composed of various automatic playlists
depending on the time of the day. Many radios also have live shows,
based on a pre-established schedule or not; a good radio software is also
expected to interrupt a live show when it becomes silent.
Most radios want to control and process the data before broadcasting
it to the public, performing tasks like volume normalization,
compression, etc.
Those examples, among many others, show the need for very flexible and modular
solutions for creating and broadcasting multimedia data. Most of the currently
available tools to broadcast multimedia data over the Internet (such as Darkice,
Ezstream, VideoLAN, Rivendell or SAM Broadcaster) consist of straightforward
adaptation of classical streaming technologies, based on
predefined interfaces, such as a virtual mixing console or static file-based
setups. Those tools perform very well a predefined task, but offer little
flexibility and are hard to adapt to new situations.

In this paper, we present \liquidsoap, a domain-specific language for
multimedia streaming.
\liquidsoap\ has established itself as one of
the major tools for audio stream generation.
The language approach has proved successful:
beyond the obvious goal of allowing the flexible combination of
common features, unsuspected possibilities have often been revealed
through clever scripts.
Finally, the modular design of \liquidsoap\ has helped its development and
maintenance, enabling the introduction of several exclusive features
over time.
\liquidsoap\ has been developed since 2004
as part of the \savonet\ project~\cite{liquidsoap}.
It is implemented in OCaml,
and we have thus also worked on interfacing many C libraries for OCaml.
The code contains approximatively $20$K lines of OCaml code
and $10$K lines of C code and runs on all major operating systems.
The software along with its documentation is freely
available~\cite{liquidsoap} under an open-source license.
%

Instead of concentrating on detailing fully the abstractions
manipulated in \liquidsoap\ (streams and sources)
or formally presenting the language and its type system,
this paper provides an overview of the two, focusing on some
key aspects of their integration.
We first give a broad
overview of the language and its underlying model in
Section~\ref{sec:liq}.
We then describe two recent extensions of that basic setup.
In Section~\ref{sec:content} we illustrate how various type system
features are combined to control the combination of stream
of various content types.
Section~\ref{sec:clocks} motivates
the interest of having multiple time flows (clocks)
in a streaming system, and presents how this feature is
integrated in \liquidsoap.
We finally discuss related systems in Section~\ref{sec:related}
before concluding in Section~\ref{sec:conclu}.

\section{Liquidsoap}
\label{sec:liq}
\subsection{Streaming model}
\label{sec:model}
A stream can be understood as a timed sequence of data.
In digital signal processing, it will simply be an infinite sequence of 
samples -- floating point values for audio, images for video.
However, multimedia streaming also involves more high-level notions.
A \emph{stream}, in \liquidsoap{}, is a succession of \emph{tracks}, annotated
with \textit{metadata}. Tracks may be finite or infinite, and can be thought
of as individual songs on a musical radio show. 
Metadata packets are punctual and can occur at any instant in the
stream. They are used to store various information about the
stream, such as the title or artist of the current track,
how loud the track should be played,
or any other custom information.
Finally, tracks contain multimedia data (audio, video or MIDI),
which we discuss in Section~\ref{sec:content}.

Streams are generated on the fly and interactively by \emph{sources}.
The behavior of sources may be affected by various parameters,
internal (\eg\ metadata) or external (\eg\ execution of commands
made available via a server).
Some sources
purely produce a stream, getting it from an external device (such as a file, a
sound card or network) or are synthesizing it.
Many other sources are actually operating on other sources in the sense that
they produce a stream based on input streams given by other sources.
Abstractly, the program describing the
generation of a stream can thus be represented by a directed acyclic graph,
whose nodes are the sources and whose edges indicate dependencies between
sources (an example is given in Figure~\ref{fig:sharing}).


Some sources have a particular status: not only do they compute a stream like any
other source, but they also perform some observable tasks, typically outputting
their stream somewhere. These are called \emph{active} sources. Stream
generation is performed ``on demand'': active sources actively attempt to
produce their stream, obtaining data from their input sources which in turn
obtain data from their dependent sources, and so on. An important consequence of
this is the fact that \emph{sources do not constantly stream}: if a source would
produce a stream which is not needed by any active source then it is actually
frozen in time.
This avoids useless computations, but is also crucial to obtain
the expected expressiveness. For example, a \texttt{rotation} operator will play
alternatively several sources, but should only rotate at the end of tracks, and
its unused sources should not keep streaming, otherwise we might find them in
the middle of a track when we come back to playing them.
\emph{Sources are also allowed to fail} after the
end of a track,
\ie{} refuse to stream, momentarily or not.
This is needed, for example,
for a queue of user requests which might often be empty,
or a playlist which may take too long to prepare a file for streaming.
Failure is handled by various operators, the most common being
the \emph{fallback}, which takes a list of sources and replays the
stream of the first available source, failing when all of them failed.

\subsection{A language for building streaming systems}

Based on the streaming model presented above, \liquidsoap{} provides the user
with a convenient high-level language for describing streaming systems.
Although our language borrows from other functional programming languages, it is
has been designed from scratch in order to be able to have a dedicated static
typing discipline together a very user-friendly language. 

One of the main goals which has motivated the design of the \liquidsoap{}
language is that it should be very accessible, even to non-programmers. It
turned out that having a functional programming language is very natural (\cf
Section~\ref{sec:transitions}). The built-in functions of the language often
have a large number of parameters, many of which have reasonable default 
values, and  it would be very cumbersome to have to write them all each time,
in the right order.
In order to address this, we have designed a new extension of
$\lambda$-calculus with labeled arguments and multi-abstractions which makes it
comfortable to use the scripting
API~\cite{baelde-mimram:webradio-lambda}. Having designed our own language also
allowed us to integrate a few domain-specific extensions, to display helpful
error messages and to generate a browsable documentation of the scripting API. In
practice, many of the users of \liquidsoap{} are motivated by creating a radio
and not very familiar with programming, so it can be considered that the design of
the language was a success from this point of view.

An other motivation was to ensure some safety properties of the stream
generation. A script in \liquidsoap{} describes a system that is intended to run
for months, some parts of whose rarely triggered, and it would be very
disappointing to notice a typo or a basic type error only after a particular
part of the code is used for an important event. In order to ensure essential
safety properties, the language is statically and strongly typed.
We want to put as
much static analysis as possible, as long as it doesn't put the burden on the
user, \ie{} all types should be inferred. As we shall see, \liquidsoap{} also
provides a few useful dynamic analysis.


\begin{figure}[t]
 \begin{center}
\[
\xymatrix{
  *+[F]{\mathtt{input.http}}\ar[r]&*+[F]{\mathtt{fallback}}\ar[r]&
  *+[F]{\mathtt{normalize}}\ar[r]\ar[dr]&
  *+<10pt>[F=:<30pt>]{\mathtt{output.icecast}}\\
  *+[F]{\mathtt{playlist}}\ar[ur]& & & *+<10pt>[F=:<30pt>]{\mathtt{output.file}}\\
}
\]
\end{center}
\fcaption{A streaming system with sharing}
\label{fig:sharing}
\end{figure}

The current paper can be read without a prior understanding of the language and
its typing system, a detailed presentation can however be found
in~\cite{baelde-mimram:webradio-lambda}.
A basic knowledge of programming languages should be enough to understand
the few examples presented in this paper,
which construct sources using built-in operators of our language.
For example, the following script defines two elementary sources, respectively 
reading from an HTTP stream and a playlist of files, composed in a fallback
and filtered through a volume normalizer.
The resulting stream is sent to an Icecast server which 
broadcasts the stream to listeners,
and saved in a local backup file:
\begin{verbatim}
s = normalize(fallback([input.http("http://other.net/radio"),
                        playlist("listing.txt")])))
output.icecast(%vorbis,mount="myradio",s)
output.file(%vorbis,"backup.mp3",s)
\end{verbatim}
The graph underlying the system resulting from the execution of that script is
shown in Figure~\ref{fig:sharing}.
Note that the two output functions build a new source:
these sources generate the same stream as \verb.s.,
but are active and have the side effect of encoding and sending
the stream, respectively to an Icecast server and a file.
A few remarks on the syntax: the notation
\hbox{\texttt{[}\ldots\texttt{]}} denotes a list, \texttt{mount}
is a label (the name of an argument of the function \texttt{output.icecast}) and
\texttt{\%vorbis} is an encoding format parameter whose meaning is explained in
Section~\ref{sec:typing-ex}
(recall that Vorbis is a compressed format for audio, similar to MP3).

%
%

\subsection{Functional transitions}
\label{sec:transitions}

\liquidsoap{} is a \emph{functional} programming language and a particularly
interesting application of this is the case of \textit{transitions}.
Instead of simply sequencing tracks, one may want a smoother transition.
For example, a \emph{crossfade} consists in mixing the end of the old source,
whose volume is faded out, with the
beginning of the new one, whose volume is faded up (see Figure~\ref{fig:cross}).
But there is a wide variety of other possible transitions:
a delay might be added, jingles may be inserted, etc.

\begin{figure}[h]
 \begin{center}
\begin{tikzpicture}[xscale=0.8,yscale=0.8]
\draw[->] (0,0) -- (0,2.5);
\draw (-0.1,2) -- (0.1,2);
\draw (0,2) node[anchor=east]{100};
\draw (0,0) node[anchor=east]{0};
\draw[->] (0,0) -- (7.5,0);
\foreach \x in {1,2,3,4,5,6,7} \draw (\x,-0.1) -- (\x,0.1);
\draw (0,2.5) node[anchor=south]{volume (\%)};
\draw (7.5,0) node[anchor=west]{time (sec)};
\draw (0,2) -- (2,2) -- (5,0);
\draw (2,0) -- (4,2) -- (7.5,2);
\draw (1,2) node[anchor=south]{old};
\draw (6,2) node[anchor=south]{new};
\end{tikzpicture}
\end{center}
 \fcaption{A crossfade transition between two tracks}
 \label{fig:cross}
\end{figure}

A solution that is both simple and flexible is to allow the user
to specify a transition as a function that combines two sources
representing the old and new tracks.
We illustrate this feature with an example involving transitions
used when switching from one source to another in a \verb.fallback..
This is particularly useful when the fallback is track insensitive,
\ie\ performs switching as soon as possible,
  without waiting for the end of a track.
The following code defines a fallback source which performs a crossfade when
switching from one source to another:
\begin{verbatim}
def crossfade(old,new) =
  add([fade.initial(duration=2.,new),fade.final(duration=3.,old)])
end
t = [crossfade,crossfade]
f = fallback(track_sensitive=false,transitions=t,[r,s])
\end{verbatim}

Because any function can be used to define a transition, the possibilities
are numerous. A striking example from the standard library of \liquidsoap{}
scripts is the operator \texttt{smooth\_add}, which takes as
argument a main source (\eg\ musical tracks) and a special interruption
source (\eg\ news items).
When a new interruption is available, \texttt{smooth\_add} gradually
reduces the volume of the main source to put it in the background,
and superposes the interruption. The reverse is performed at the
end of the interruption.
This very appreciated effect is programmed using the same building blocks
as in the previous example.

\subsection{Efficient implementation}

An important aspect of the implementation is efficiency concerning both CPU and
memory usage. The streams manipulated can have high data rates (a typical video
stream needs 30Mo/s) and avoiding multiple copies of stream data is
crucial.

In \liquidsoap, streams are computed using \emph{frames},
which are data buffers representing a portion of stream portion of fixed
duration.
Abstractly, sources produce a stream by producing a sequence of frames.
However, in the implementation a source is \emph{passed} a frame
that it has to fill.
Thus, we avoid unnecessary copies and memory allocations.
Active sources, which are the initiators of streaming,
initially allocate one frame, and keep re-using it to get stream data
from their input source.
Then, most sources do not need to allocate their own frame,
they simply pass frames along and modify their content in place.
However, this simple mechanism does not work when a source is \emph{shared},
\ie\ it is the input source of several sources.
This is the case of the \verb.normalize. node in the graph of
Figure~\ref{fig:sharing} (which happens to be shared by active sources).
In that case, we use a \emph{caching} mechanism:
the source will have its own \emph{cache} frame
for storing its current output.
The first time that the source is asked to fill a frame,
it fills its internal cache and copies the data from it;
in subsequent calls it simply fills the frame with the data computed
during the first call.
Once all the filling operations have been done, the
sources are informed that the stream has moved on to the next frame and can
forget their cache.

With this system, frames are created once for all, one for each active source
plus one for each source that is shared and should thus perform caching |
of course, some sources might also need another frame
depending on their behavior.
Sharing is detected automatically when the source is initialized for
streaming. We do not detail this analysis, but note that the dynamic
reconfigurations of the streaming system (notably caused by transitions)
make it non-trivial to anticipate all possible sharing situations without
over-approximating too much.


\section{Heterogeneous stream contents}
\label{sec:content}



In \liquidsoap{}, streams can contain data of various nature. The typical
example is the case of video streams which usually contain both images and
audio samples. We also support MIDI streams (which contain musical notes)
and it would be easy to add other kinds of content.
It is desirable to allow sources of different
content kinds within a streaming system, which makes it necessary to
introduce a typing discipline in order to ensure 
the consistency of stream contents across sources.

The nature of data in streams is described by its \emph{content type}, which is
a triple of natural numbers indicating the number of audio, video and midi
channels.  A stream may not always contain data of the same type.  For instance,
the \texttt{playlist} operator might rely on decoding files of heterogeneous
content, \eg\ mono and stereo audio files.  In order to specify how content
types are allowed to change over time in a stream, we use \emph{arities}, which
are essentially natural numbers extended with a special symbol $\star$:
\[
a\quad ::=\quad \star \;|\; 0 \;|\; S(a)
\]
An arity is \emph{variable} if it contains $\star$, otherwise it is an usual
natural number, and is \emph{fixed}. A \emph{content kind} is a triple of
arities, and specifies which content types are acceptable. For example,
$(S(S(0)),S(\star),\star)$ is the content kind meaning ``2 audio channels, at
least one video channel and any number of MIDI channels''.
This is formalized through the subtyping relation defined in
Figure~\ref{fig:subtyping}: $T\sub K$ means
that the content kind $T$ is allowed by $K$. More generally,
$K \sub K'$ expresses that $K$ is more permissive than $K'$,
which implies that a source of content kind $K$ can safely be seen
as one of content kind $K'$.

\begin{figure}[htpb]\[
   \infer{0\sub 0}{} \quad\quad
   \infer{S(A)\sub S(A')}{A\sub A'} \quad\quad
   \infer{\star\sub\star}{} \quad\quad
   \infer{0\sub \star}{} \quad\quad
   \infer{S(A)\sub \star}{A\sub\star}
\]\[
   \infer{(A,B,C) \sub (A',B',C')}{A \sub A' & B\sub B' & C\sub C'}
\]
 \fcaption{Subtyping relation on arities}
 \label{fig:subtyping}
\end{figure}

When created, sources are given their expected content kind.
Of course, some assignments are invalid.
For example,
a pure audio source cannot accept a content kind which requires video 
channels, and many operators cannot produce a stream of an other kind
than that of their input source.
Also, some sources have to operate on input streams that have a fixed kind --
a kind is said to be fixed when all of its components are.
This is the case of the \texttt{echo} operator which produces echo on sound
and has a internal buffer of a fixed format for storing past sound,
or sound card inputs/outputs which have to initialize the sound card for
a specific number of channels.
Also note that passing the expected content kind
is important because some sources behave differently depending on their kind,
as shown with the previous example.

\label{sec:typing-ex}
\begin{figure}[b]
  \centering
  \texttt{
    \begin{tabular}{rcl}
    swap&:&(source(2,0,0)) -> source(2,0,0)\\
    on\_metadata&:&(handler,source('*a,'*b,'*c)) -> source('*a,'*b,'*c)\\
    echo&:&(delay:float,source('\#a,0,0)) -> source('\#a,0,0)\\
    greyscale&:&(source('*a,'*b+1,'*c)) -> source('*a,'*b+1,'*c)\\
    output.file &:& 
       (format('*a,'*b,'*c),string,source('*a,'*b,'*c))->\\
     & & source('*a,'*b,'*c) \\
     & &
    \end{tabular}
  }
  \fcaption{Types for some operators}
  \label{fig:types}
\end{figure}

\paragraph{Integration in the language.}
To ensure that streaming systems built from user scripts will never
encounter situations where a source receives data that it cannot handle,
we leverage various features of our type system.
By doing so, we guarantee statically that content type mismatches never happen.
The content kinds are reflected into types,
and used as parameters of the \texttt{source} type.
In order to express the types of our various operators,
we use a couple features of type systems
(see \cite{pierce02book} for extensive details).
As expected, the above subtyping relation is integrated into
the subtyping on arbitrary \liquidsoap\ types.
We illustrate various content kinds in the examples of Figure~\ref{fig:types}:
\begin{itemize}
\item The operator \texttt{swap} exchanges the two channels of a stereo audio
  stream. Its type is quite straightforward: it operates on streams with exactly
  two audio channels.
\item
  \liquidsoap\ supports polymorphism \emph{à la} ML.
  We use it in combination with constraints to allow arbitrary arities.
  The notation \verb.'*a. stands for a universal variable (denoted
  by \verb.'a.) to which a type constraint is attached, expressing that
  it should only be instantiated with arities.
  For example, the operator \verb.on_metadata. does not rely
  at all on the content of the stream, since it is simply in
  charge of calling a handler on each of its metadata packets --
  in the figure, \verb.handler. is a shortcut for
  \verb.([string*string]) -> unit..
\item When an operator, such as \verb.echo., requires a fixed content type, we
  use another type constraint. The resulting constrained universal
  variable is denoted by \verb.'#a. and can only be instantiated with
  fixed arities.
\item The case of the \texttt{greyscale} operator, which converts a color
  video into greyscale, shows how we can require at least one video channel in
  types.  Here, \verb.'*b+1. is simply a notation for \verb.S('*b)..
\item Finally, the case of \verb#output.file# (as well as several other outputs
  which encode their data before sending it to various media) is quite
  interesting. Here, the expected content kind depends on the format the stream
  is being encoded to, which is given as first argument of the operator. Since
  typing the functions generating formats would require dependent types (the
  number of channels would be given as argument) and break type inference, we
  have introduced particular constants for type formats with syntactic sugar for
  them to appear like functions -- similar ideas are for example used to type
  the \texttt{printf} function in OCaml. For example,
  \verb$output.file(%vorbis,"stereo.ogg",s)$ requires that
  \verb.s. has type \verb.source(2,0,0). because \verb.%vorbis. alone
  has type \verb.format(2,0,0).,
  but \verb$output.file(%vorbis(channels=1),"mono.ogg",s)$
  requires that there is only one audio channel;
  we also have video formats such as \verb.%theora..
\end{itemize}

These advanced features of the type system are statically inferred, which means
that the gain in safety does not add any burden on users. As said above, content
kinds have an influence on the behavior of sources |
polymorphism is said to be \emph{non-parametric}.
In practice, this means that static types must be
maintained throughout the execution of a script. This rather unusual aspect
serves us as an overloading mechanism: the only way to remove content kinds from
execution would be to duplicate our current collection of operators with a
different one for each possible type instantiation.


~

Ideally, we would like to add some more properties to be statically checked by
typing. But it is sometimes difficult to enrich the type system while keeping
a natural syntax and the ability to infer types.
For example,
\liquidsoap{} checks that active sources are \emph{infallible}, \ie{} always
have data available in their input stream, and this check is currently done by a
flow analysis on instantiated sources and not typing.
Another example is clocks which are described next
section.

\section{Clocks}
\label{sec:clocks}


Up to now, we have only described streaming systems where there is
a unique global clock. In such systems, time flows at the same rate
for all sources.
By default, this rate corresponds to the wallclock time,
which is appropriate for a live broadcast,
but it does not need to be so.
For example, when producing a file from other files,
one might want the time rate to be as fast as the CPU allows.


While having a global clock suffices in many situations,
there are a couple of reasons why a streaming system might involve multiple
clocks or time flows.
The first reason is external to liquidsoap: there is simply
not a unique notion of time in the real world.
A computer's internal clock indicates a slightly different time
than your watch or another computer's clock.
Moreover, when communicating with a remote computer, network
latency causes a perceived time distortion.
Even within a single computer there are several clocks: notably, each
soundcard has its own clock, which will tick at a slightly different
rate than the main clock of the computer.
Since liquidsoap communicates with soundcards and remote computers,
it has to take those mismatches into account.

There are also some reasons that are purely internal to liquidsoap:
in order to produce a stream at a given speed,
a source might need to obtain data from another source at
a different rate. This is obvious for an operator that speeds up or
slows down audio, but is also needed in more subtle cases
such as a crossfading operator. A variant of the operator described
in Section~\ref{sec:transitions} might combine a portion
of the end of a track with the beginning of the next track
\emph{of the same source} to create a transition between tracks.
During the lapse of time where the operator combines data from an end of track
with the beginning of the other other, the crossing operator needs to read both
the stream data of the current track and the data of the next track, thus
reading twice as much stream data as in normal time.
After ten tracks,
with a crossing duration of six seconds, one more minute will have
passed for the source compared to the time of the crossing operator.

\subsection{Model}

In order to avoid inconsistencies caused by time differences,
while maintaining a simple and efficient execution model for
its sources, liquidsoap works under the restriction that
one source belongs to a unique clock,
fixed once for all when the source is created.
Sources from different clocks cannot communicate using the normal
streaming protocol, since it is organized around clock cycles:
each clock is responsible for animating its own active sources
and has full control on how it does it.

In the graphical representation of streaming systems,
clocks induce a partition of sources represented by a notion of locality
or box, and clock dependencies are represented by nesting.
For example, the graph shown in Figure~\ref{fig:boxes}
corresponds to the stream generators built by the following
script:
\begin{verbatim}
output.icecast(%vorbis,mount="myradio",
  fallback([crossfade(playlist("some.txt")),jingles]))
\end{verbatim}

\begin{figure}[t]
 \begin{center}
\[
\def\f{\save
*+<15pt>[F--]\frm{}\ar @{--} "2,2"\restore}%
\def\g{\save
"2,4"."1,2"."1,5"!C*+<27pt>[F--]\frm{}\ar @{--} "1,1"\restore}%
\xymatrix{
   {clock_1} & *+[F]{\mathtt{playlist}}\ar[r]\f&*+[F]{\mathtt{crossfade}}\ar[r]&  *+[F]{\mathtt{fallback}}\ar[r]&
  *+[F]{\mathtt{output.icecast}}\\
   &{clock_2} &  & *+[F]{\mathtt{jingles}}\ar[u]\g& 
}
\]
\end{center}
 \fcaption{A streaming system with two clocks}
 \label{fig:boxes}
\end{figure}

There, $clock_2$
was created specifically for the crossfading
operator; the rate of that clock is controlled by that operator,
which can hence accelerate it around track changes without any
risk of inconsistency.
$clock_1$ is simply a wallclock, so that the main stream
is produced following the real time rate.

A clock is \emph{active} if it ticks by itself,
therefore running its sources constantly; this
is the case of wallclocks or soundcard clocks.
We say that a clock depends on another one
if its animation (and thus time rate) depends on it.
Active sources do not depend on other sources,
and dependencies must be acyclic.
In the above example, the ticking of
$clock_2$ is provoked by that of
$clock_1$, and freezes when the fallback
is playing jingles.
Although nothing forces it in the model, it makes more sense if
each passive source depends (possibly indirectly) on an active one,
and all sources without dependencies are active.
Those assumptions are in fact guaranteed to hold for the systems
built using the \liquidsoap{} language.

From an implementation viewpoint, each active clock launches
its own streaming thread.
Hence, clocks provide a way to split the generation of one or
several streams across several threads,
and hence multiple CPU cores.
This powerful possibility is made available to the user
through the intuitive notion of clock.
As we shall see in the next section,
the script writer never needs to specify clocks unless he
explicitly wants a particular setup,
and \liquidsoap\ automatically checks that clock assignements
are correct.

\subsection{Clock assignment}

Clocks are not represented in the type of \liquidsoap\ sources.
Although it would be nice to statically check clock assignment,
type inference would not be possible without technical annotations
from the user. Instead, clocks are assigned upon source creation.
Some sources require to belong to a particular, definite clock,
such as the wallclock, or the clock corresponding to a sound card.
Most sources simply require that their clock is the same as their
input sources.
Since clocks often cannot be inferred bottom-up, we use a notion
of clock variable that can be left undefined.
Clock variables reflect the required clock dependencies,
which are maintained during the inference process.

Two errors can occur during this phase.
Although they are runtime errors that could be raised
in the middle of streaming when new sources are created
(\eg{} by means of a transition),
this usually only happens during the initial construction.
The first error is raised when
two different known clocks need to be unified.
For example, in the following script, the ALSA input is
required to belong to the ALSA clock and \verb.crossfade.'s internal clock
at the same time:
\begin{verbatim}
output.file(%vorbis,"record.ogg",crossfade(input.alsa()))
\end{verbatim}
The other possible error happens when unifying two unknown clock variables
if one depends on the other -- in unification terminology, this is an
\emph{occurs-check} failure. A simple example of that situation is
the script \verb.add([s,crossfade(s)]). where the two mixed sources
respectively have clocks $c$ and $X_c$ where $c$ is the clock created
for the crossfading operator and $X_c$ is the variable representing
the clock to which the crossfading belongs, on which $c$ depends.

After this inference phase, it is possible that some clocks are still
unknown. Remaining variables are thus forcibly assigned to the default
wallclock, before that all new sources are prepared for streaming
by their respective clocks.

\section{Related work} \label{sec:related}

\liquidsoap{} is obviously different from classical tools
such as Ices or Darkice in
that it offers the user the freedom to assemble a stream
for a variety of operators, through a scripting
language rather than traditional configuration files.

\liquidsoap{} has more similarities with multimedia streaming libraries
and digital signal processing (DSP) languages.
The GStreamer library~\cite{gstreamer} defines a model of stream, and its API
can be used to define streaming systems in various programming
languages (primarily coded in C, the library has also been
ported to many other languages).
Faust~\cite{faust} provides a high-level functional programming language for
describing stream processing devices,
and compiles this language down to C++, which enables an integration
with various other systems.
It is also worth mentioning Chuck~\cite{chuck}, a DSP programming
language with an emphasis on live coding (dynamic code update).
Besides a different approach and target application,
\liquidsoap{} differs more deeply from these tools.
The notion of source provides a richer way of
generating streams, providing and relying on the additional notions
of tracks and metadata; also recall the ability to momentarily
stop streaming, and the possibility to dynamically create or
destroy sources.
It would be very interesting to interface \liquidsoap{} with the
above mentionned tools, or import some of their techniques.
This could certainly be done for simple operators such as DSP,
and would allow us to program them efficiently and declaratively
from the scripting language rather than in OCaml.

\section{Conclusion} \label{sec:conclu}

We have presented the main ideas behind the design of \liquidsoap,
a tool used by many netradios worldwide as well as in some academic 
work~\cite{baccigalupo2007case}.
We believe that \liquidsoap\ demonstrates the potential of
building applications as domain-specific languages.
It also shows that very rich type systems can be put to work
usefully even in tools not designed for programmers:
although most \liquidsoap\ users have a limited understanding of
our type system, they are able to fix their mistakes
when an error is reported | errors might be difficult to read but
they have the merit of signaling real problems.

Of course, there are many other reasons behind the success of \liquidsoap,
including a wide variety of features plugged onto the basic organization
described here. 
Some of the future work on \liquidsoap\ lies there:
integration with other tools, graphical interfaces, documentation, etc.
But we are also planning some improvements of the language.
One of the goals is to make it possible to express more operators
directly in \liquidsoap\ instead of OCaml,
bringing more customizability to the users.
Also, \liquidsoap\ offers a server through which many sources
offer various services. An interesting way to structure more this very useful
system would be to consider sources as objects whose methods are services,
and type them accordingly.

\bibliographystyle{abbrv}
\bibliography{biblio}
\end{document}